\documentclass[final,5p,times,twocolumn]{elsarticle}

\usepackage[T1]{fontenc}
\usepackage[utf8]{inputenc}

\usepackage{amsmath}
\usepackage{amsfonts}
\usepackage{amssymb}
\usepackage{amsxtra}
\usepackage{array}
\usepackage{color}
\usepackage{dcolumn}
\usepackage{graphicx}
\usepackage{hepunits}
\usepackage{xspace}
\usepackage{CJKutf8}

\definecolor{purple}{rgb}{0.5,0,0.5}
\definecolor{blue}{rgb}{0.0,0,0.9}
\definecolor{prdblue}{rgb}{0.133,0.118,0.498}
\usepackage[colorlinks=true, pdfstartview=FitV, linkcolor=prdblue, citecolor= prdblue, urlcolor=prdblue]{hyperref}

\usepackage[mathscr,scaled=1.15]{urwchancal}
\DeclareFontFamily{OT1}{pzc}{}
\DeclareFontShape{OT1}{pzc}{m}{it}%
{<-> s * [1.15] pzcmi7t}{}
\DeclareMathAlphabet{\mathpzc}{OT1}{pzc}{m}{it}

\biboptions{sort&compress}

\journal{Physics Letters B}

\newcommand{\beq}{\begin{equation}}
\newcommand{\eeq}{\end{equation}}
\newcommand{\ba}{\begin{array}}
\newcommand{\ea}{\end{array}}
\newcommand{\bea}{\begin{align}}
\newcommand{\eea}{\end{align}}
\newcommand{\bi}{\begin{itemize}}
\newcommand{\ei}{\end{itemize}}
\newcommand{\ben}{\begin{enumerate}}
\newcommand{\een}{\end{enumerate}}
\newcommand{\bc}{\begin{center}}
\newcommand{\ec}{\end{center}}
\newcommand{\bl}{\begin{flushleft}}
\newcommand{\el}{\end{flushleft}}
\newcommand{\br}{\begin{flushright}}
\newcommand{\er}{\end{flushright}}

\begin{document}
\begin{CJK*}{UTF8}{gbsn}


\begin{frontmatter}

\title{$\,$\\[-7ex]\hspace*{\fill}{\normalsize{\sf\emph{Preprint no}. NJU-INP 095/24}}\\[1ex]
%
Likelihood of a zero in the proton elastic electric form factor}

\author[AHNU]{Peng Cheng (程鹏)%
    $\,^{\href{https://orcid.org/0000-0002-6410-9465}{\textcolor[rgb]{0.00,1.00,0.00}{\sf ID}}}$}

\author[UHe,UPO]{Zhao-Qian Yao (姚照千)%
       $\,^{\href{https://orcid.org/0000-0002-9621-6994}{\textcolor[rgb]{0.00,1.00,0.00}{\sf ID}}}$}

\author[ECT]{Daniele Binosi%
    $\,^{\href{https://orcid.org/0000-0003-1742-4689}{\textcolor[rgb]{0.00,1.00,0.00}{\sf ID}}}$}

\author[NJU,INP]{Craig D. Roberts%
       $^{\href{https://orcid.org/0000-0002-2937-1361}{\textcolor[rgb]{0.00,1.00,0.00}{\sf ID}},}$}

\address[AHNU]{Department of Physics, \href{https://ror.org/05fsfvw79}{Anhui Normal University}, Wuhu, Anhui 24100, China}

\address[UHe]{Department of Integrated Sciences and Center for Advanced Studies in Physics, Mathematics and Computation, \href{https://ror.org/03a1kt624}{University of Huelva}, E-21071 Huelva, Spain}

\address[UPO]{Dpto. Sistemas F\'isicos, Qu\'imicos y Naturales, Univ.\ Pablo de Olavide, E-41013 Sevilla, Spain}

\address[ECT]{European Centre for Theoretical Studies in Nuclear Physics
            and Related Areas  (\href{https://ror.org/01gzye136}{ECT*}), Villa Tambosi, Strada delle Tabarelle 286, I-38123 Villazzano (TN), Italy}

\address[NJU]{
School of Physics, \href{https://ror.org/01rxvg760}{Nanjing University}, Nanjing, Jiangsu 210093, China}
\address[INP]{
Institute for Nonperturbative Physics, \href{https://ror.org/01rxvg760}{Nanjing University}, Nanjing, Jiangsu 210093, China\\[1ex]
%
\href{mailto:binosi@ectstar.eu}{binosi@ectstar.eu} (DB);
\href{mailto:cdroberts@nju.edu.cn}{cdroberts@nju.edu.cn} (CDR)
\\[1ex]
Date: 2024 December 13\\[-6ex]
}

\begin{abstract}
Working with the $29$ available data on the ratio of proton electric and magnetic form factors, $\mu_p G_E^p(Q^2)/ G_M^p(Q^2)$, and independent of any model or theory of strong interactions, we use the Schlessinger point method to objectively address the question of whether the ratio possesses a zero and, if so, its location.  Our analysis predicts that, with 50\% confidence, the data are consistent with the existence of a zero in the ratio on $Q^2 \leq 10.37\,$GeV$^2$.  The level of confidence increases to $99.9$\% on $Q^2 \leq 13.06\,$GeV$^2$.  Significantly, the likelihood that existing data are consistent with the absence of a zero in the ratio on $Q^2 \leq 14.49\,$GeV$^2$ is $1/1$-million.
\end{abstract}


\end{frontmatter}
\end{CJK*}

\section{Introduction}
\label{s1}
The proton is Nature's most fundamental composite system.  It was discovered over one hundred years ago and, during the intervening years, a great deal of empirical information has been gathered on proton properties.  Today, by any reasonable measure, it may be considered stable: the lower limit on the proton lifetime is reckoned to be more than $10^{23}$-times the age of the Universe; and its mass, $m_p$, seems to set a basic scale for life and matter.

Within the Standard Model of particle physics, the proton is supposed to be constituted from three light valence quarks $u + u +d$, bound by interactions described by quantum chromodynamics (QCD).  However, despite QCD itself being a fifty-year-old theory, a solution to the proton bound state problem has not yet been found.  Approximate treatments and numerical studies exist; yet, many questions remain unanswered.  They may all be collected under a single umbrella, \emph{viz}.\ how do the mass, spin, and structure of the proton emerge from interactions between the (practically) massless degrees-of-freedom used to express the QCD Lagrangian density?  Perspectives on these problems may be found elsewhere \cite{Roberts:2021nhw, Binosi:2022djx, Ding:2022ows, Roberts:2022rxm, Ferreira:2023fva, Carman:2023zke, Salme:2022eoy, deTeramond:2022zcm, Krein:2023azg}.

Throughout the history of proton structure investigations, electron scattering has played a key role \cite{Punjabi:2015bba, Gao:2021sml}.  In elastic electron + proton scattering, the proton contribution to the associated matrix element may be described by the following current:
\begin{align}
J_\mu^p(Q) & = ie \Lambda_+(p_f)
[ F_1^p(Q^2) \gamma_\mu \nonumber \\
& \quad + \frac{1}{2 m_p} \sigma_{\mu\nu} Q_\nu F_2^p(Q^2) ]
\Lambda_+(p_i)\,,  \label{NucleonCurrent}
\end{align}
where $e$ is the positron charge,
$m_p$ is the proton mass,
the incoming and outgoing proton momenta are $p_{i,f}$, $Q=p_f-p_i$ is the momentum transfer to the proton in the interaction,
$\Lambda_+(p_{i,f})$ are positive-energy proton-spinor projection operators,
and $F_{1,2}^p$ are the Dirac and Pauli form factors.
The interaction current can equally be expressed in terms of the proton charge and magnetisation distributions ($\tau = Q^2/[4 m_N^2]$) \cite{Sachs:1962zzc}:
\begin{equation}
G_E^p  = F_1^p - \tau F_2^p\,,
\quad G_M^p  = F_1^p + F_2^p \,.
\label{Sachs}
\end{equation}

In the last century, it was suggested that, with $\mu_p = G_M^p(0)$, $\mu_p G_E^p(Q^2)/ G_M^p(Q^2) \approx\,$ constant, \emph{i.e}., this ratio is independent of momentum transfer, so that proton electric charge and magnetisation distributions are effectively identical.  This was a data-driven conclusion \cite{Litt:1969my, Berger:1971kr, Price:1971zk, Bartel:1973rf, Walker:1993vj, Andivahis:1994rq, BatesFPP:1997rpw, Sill:1992qw}.
Even so, from a quantum mechanics perspective, the suggestion could be seen as surprising: form factors computed from a wave function expressing a rigid sphere (radius $R$) bound state typically exhibit a series of diffraction zeros, with the first located at $Q^2 \approx (3 \pi/2R)^2$, \emph{i.e}.,  $\approx (1.1\,{\rm GeV})^2$ for the proton.  The zeros in $G_{E,M}^p$ need not coincide.  Only a point particle has positive definite (constant) form factors.

\begin{figure}[t]
\centerline{%
\includegraphics[clip, width=0.95\linewidth]{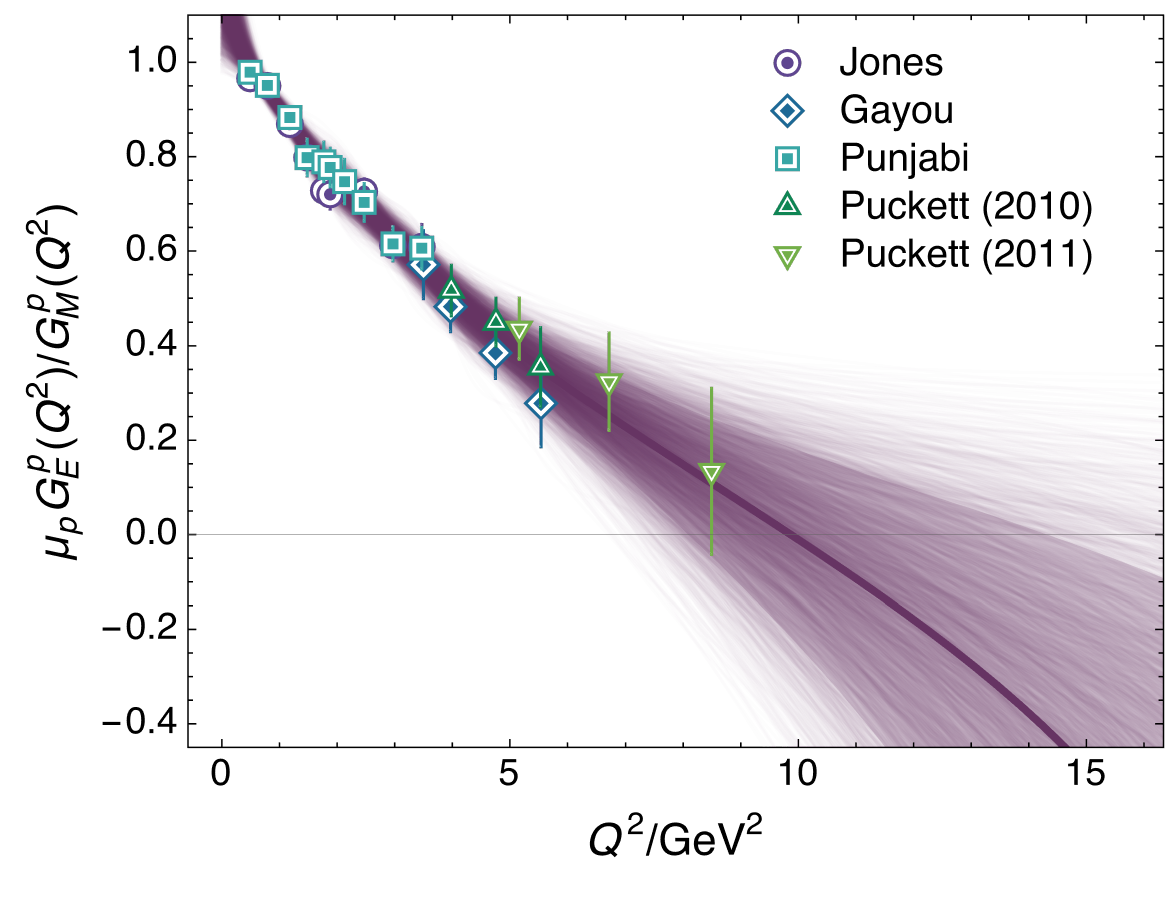}}
\caption{\label{FigData}
$\mu_p G_E^p(Q^2)/ G_M^p(Q^2)$.
Available data obtained using the polarisation transfer reaction \cite{Jones:1999rz, Gayou:2001qd, Punjabi:2005wq, Puckett:2010ac, Puckett:2017flj};
and $M=9$ analysis of this data, showing the $5\,000$ data-replica interpolating functions (purple curves) and their mean (thicker purple curve) obtained as discussed in Steps~1\,--\,4.
%
%
}
\end{figure}

The results for $G_{E,M}^p(Q^2)$ that are reported in Refs.\,\cite{Litt:1969my, Berger:1971kr, Price:1971zk, Bartel:1973rf, Walker:1993vj, Andivahis:1994rq, BatesFPP:1997rpw, Sill:1992qw} were obtained via Rosenbluth separation of $e + p \to e +p$ cross-section data.  In that approach, sensitivity to the electric form factor diminishes rapidly with increasing $Q^2$, so the magnetic form factor becomes dominant.
On the other hand, the polarisation transfer reaction $\vec{e} + p \to e + \vec{p}$ can be used to obtain a data ratio that is directly sensitive to $\mu_p G_E^p(Q^2)/ G_M^p(Q^2)$ \cite{Arnold:1980zj}, thereby avoiding issues with Rosenbluth separation.
Following the beginning of operations with the high-luminosity electron accelerator at JLab, 
such data could be obtained; and they dramatically changed the picture \cite{Jones:1999rz}.
The data indicate that $\mu_p G_E^p(Q^2)/ G_M^p(Q^2)$ falls steadily away from unity with increasing $Q^2$, reaching a value of $\approx 0.6$ at $Q^2\approx 3\,$GeV$^2$.
Additional such data have since been accumulated: the set now contains $29$ points, reaching to $Q^2 = 8.5\,$GeV$^2$ \cite{Jones:1999rz, Gayou:2001qd, Punjabi:2005wq, Puckett:2010ac, Puckett:2017flj}.
They are displayed in Fig.\,\ref{FigData}.
Evidently, the downward trend continues.
It is anticipated that the foreseeable future will deliver data on the ratio out to $Q^2=12\,$GeV$^2$ \cite{Wojtsekhowski:2014vua, Gilfoyle:2018xsa}.

Following publication of the first $\mu_p G_E^p(Q^2)/ G_M^p(Q^2)$ data and its subsequent confirmation, numerous model and theory calculations have been performed.
With varying degrees of parameter dependence and tuning, some produce a form factor ratio that is consistent with available measurements.
Amongst these are analyses that predict a zero in the ratio beyond the range of extant data -- see, \emph{e.g}., Refs.\,\cite{deMelo:2008rj, Cui:2020rmu, Yao:2024uej}.
However, not all calculations produce such a zero -- see, \emph{e.g}., Refs.\,\cite{Anikin:2013aka, Giannini:2015zia, Sufian:2016hwn, Kallidonis:2018cas, Xu:2021wwj}.

It is worth noting that relativistic effects are important in describing available $\mu_p G_E^p(Q^2)/ G_M^p(Q^2)$ data, but particular features of QCD may be equally or more significant.
For instance, owing to dynamical chiral symmetry breaking, a corollary of emergent hadron mass (EHM) \cite{Roberts:2021nhw, Binosi:2022djx, Ding:2022ows, Roberts:2022rxm, Ferreira:2023fva, Carman:2023zke}, light quarks acquire a strongly momentum dependent running mass that is large at infrared momenta \cite{Binosi:2016wcx}: $M(k^2=0) \approx 0.35\,$GeV.
Quark + interacting-diquark Faddeev equation models of proton structure \cite{Barabanov:2020jvn} suggest that the rate at which $M(k^2)$ runs toward its ultraviolet ($k^2/m_p^2 \gg 1$) current-mass limit has a material influence on the proton Pauli form factor \cite{Wilson:2011aa, Cloet:2013gva}: if the evolution is very rapid, \emph{i.e}., perturbative physics is quickly recovered, then $\mu_p G_E^p(Q^2)/ G_M^p(Q^2)$ does not exhibit a zero, whereas a zero is found with a slower transition from the nonperturbative to the perturbative domain.
Plainly, delivering a QCD explanation of the puzzling behaviour of $\mu_p G_E^p(Q^2)/ G_M^p(Q^2)$ is a high priority.


\section{Objective Analysis of Available Data}
\label{s2}
A key point is whether or not $\mu_p G_E^p(Q^2)/ G_M^p(Q^2)$ possesses a zero.  Eschewing any theory of hadron structure, we address this question by employing the Schlessinger point method (SPM) \cite{Schlessinger:1966zz, PhysRev.167.1411, Tripolt:2016cya}, to which we now provide a little background.

Suppose one has $N$ pairs, ${\mathsf D} = \{(x_i,y_i=f(x_i))$\}, being the values of some smooth function, $f(x)$, at a set of discrete points, an SPM application constructs a continued-fraction interpolation:
\begin{equation}
\label{SPMinterpolator}
{\mathpzc C}_N(x) = \frac{y_1}{1+\frac{a_1(x-x_1)}{{1+\frac{a_2(x-x_2)}{\vdots a_{N-1}(x-x_{N-1})}}}}\,,
\end{equation}
wherein the coefficients $\{a_i|i=1,\ldots, N-1\}$ are constructed recursively and ensure ${\mathpzc C}_N(x_i) = f(x_i)$, $i=1\,\ldots,N$.
The SPM may also be described as a multipoint Pad\'e approximant.
The procedure reliably reconstructs any analytic function within a radius of convergence determined by that one of the function's branch points which is closest to the domain of real-axis points containing the data sample.
For instance, given a monopole form factor represented by $N>0$ points, each one lying on the curve; then using any one of those points, the SPM will precisely reproduce the function.

Modern SPM implementations include a statistical element, so that representations of an unknown underlying curve come with a reliable quantitative estimate of uncertainty.
Crucially, the SPM is free from practitioner bias -- it is blind to any and all prejudice concerning the target function; hence, the SPM delivers objective estimates of the analytic continuations of the function being sought.

In practice, the SPM has been blind-tested against numerous models and physically validated in applications that include
extraction of hadron and light nucleus radii from electron scattering \cite{Cui:2022fyr};
inference of resonance properties from scattering data \cite{Binosi:2022ydc};
validating evidence of the odderon in high-energy elastic hadron+hadron scattering \cite{Cui:2022dcm};
and
prediction of meson and baryon electromagnetic and gravitational form factors \cite{Yao:2024drm, Yao:2024uej, Yao:2024ixu}.

\begin{figure*}[t]
\centerline{%
\includegraphics[clip, width=0.9\textwidth]{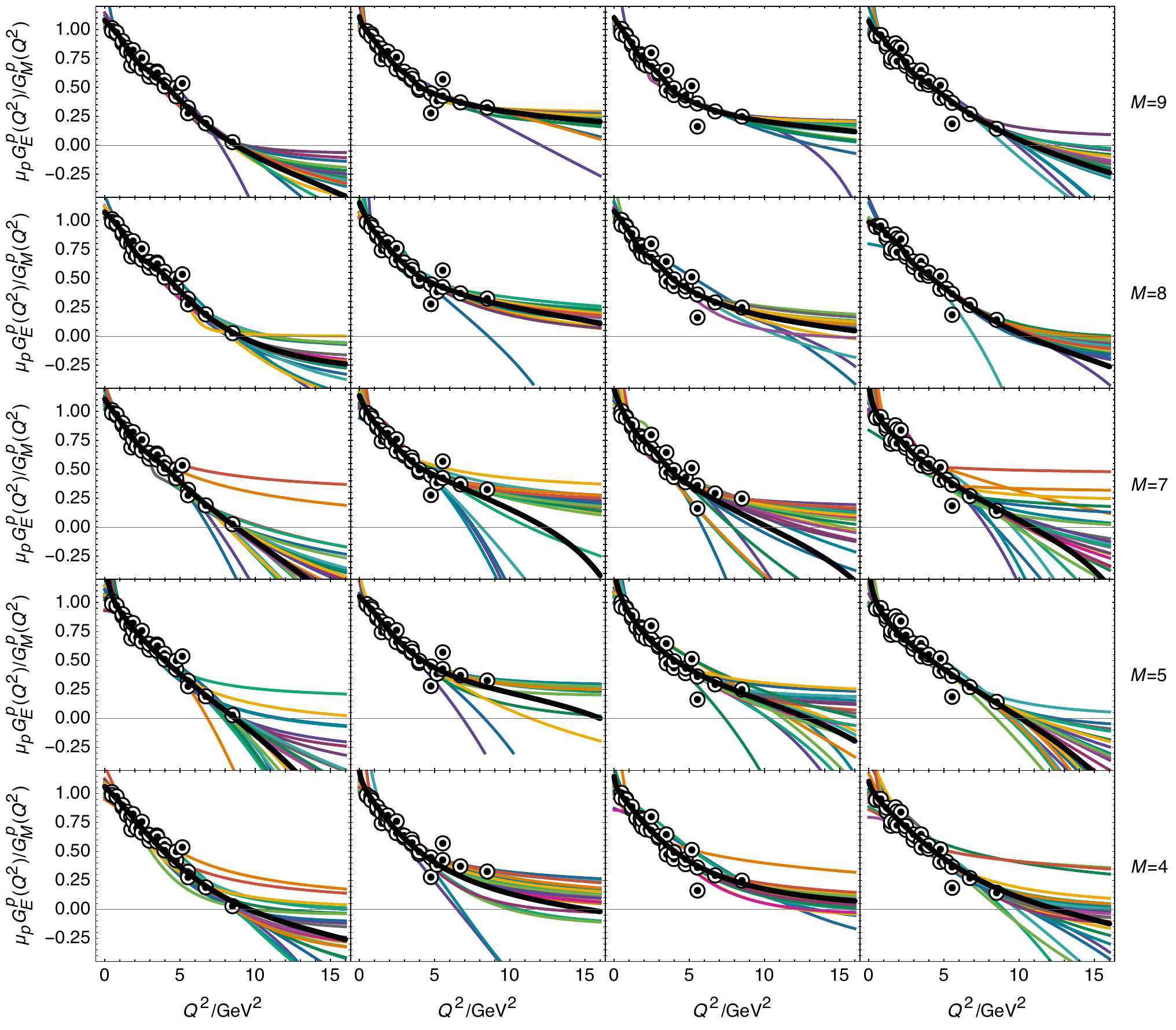}}
\caption{\label{FigSPM9}
Each row shows the same $4$ distinct, randomly selected replicas from our $5\,000$ member set.
Then, as labelled from top to bottom, the images display the $25$ associated $M$-point SPM interpolators and the averaged result (thick black curve) which is our final $M$-point SPM representative of the replica in the panel.
}
\end{figure*}


Herein, we proceed as follows.
\begin{description}
\item [Step 1] Regarding $R^p(Q^2) = \mu_p G_E^p(Q^2)/ G_M^p(Q^2)$, $N=29$ pairs are available.
As experimental data, supposing that systematic errors are small -- as is expected when polarisation transfer data are collected in simultaneous measurements of both recoil polarisation components \cite{Puckett:2017flj}, then they are distributed statistically around the underlying analytic target curve.
To accommodate this, we generate a replica of the data set by replacing each datum by a new point,  relocated according to a Gaussian distribution whose mean is the original experimental value and whose width is fixed by the associated experimental uncertainty.
We do this $15\,000$ times.

\item[Step 2]
    Taking a particular replica, we choose $M=9$ points at random and produce an interpolator in the form of Eq.\,\eqref{SPMinterpolator}.  (There are more than $3.7$-million possible choices of distinct $9$-point sets.)
    Such an interpolator is accepted into our function space so long as it is (\emph{i}) a $C^1$ function on $0<Q^2/{\rm GeV}^2<20$ and (\emph{ii}) monotonically decreasing.
    This step is repeated until we obtain $25$ acceptable interpolators on the given replica.

\item[Step 3]
    If $25$ such functions cannot be obtained, we discard the replica as being too noisy, and randomly select another from the $15\,000$ member set.
    On this new replica, we again seek $25$ interpolators.
    If successful, then we keep that replica.  If not, a new replica is chosen, etc.

\item[Step 4]
    Steps~2, 3 are repeated until we arrive at $5\,000$ independent replicas which support $25$ acceptable interpolators.
    On each of these replicas, the average of the $25$ interpolators is identified as the curve representative of the selected replica.

    \emph{N.B}. Given the large data uncertainties, especially on $Q^2 \gtrsim 3\,$GeV$^2$, this averaging procedure produces a more faithful reproduction of a given replica's content than a single interpolator -- see Fig.\,\ref{FigSPM9}.

\end{description}

Constructed as described above, the $5\,000$ $M=9$ interpolators that represent the $N=29$ available $R^p(Q^2)$ data and their uncertainties are drawn in Fig.\,\ref{FigData} -- see the purple curves.  In the process of averaging over these curves (to obtain the thick purple curve), we find that $92$\% of the individual interpolators have crossed zero on $Q^2<15\,$GeV$^2$.  The conclusion of this $M=9$ analysis is that a function-form unbiased extrapolation of available data, which accounts for its uncertainties, indicates that $R^p(Q^2)$ should exhibit a zero at the location indicated in Table~\ref{TabZero}.


We next repeat Steps 2, 4 for $M=8, 7, 5, 4$.  In these cases, the $5\,000$ replicas that support $25$ $M=9$ SPM interpolators also carry $25$ interpolators for the other $M$ values.  So, Step 3 is unnecessary and we have the same set of $5\,000$ replicas for each value of $M$.
In all cases, the vast majority of interpolators predict a zero in $R^p(Q^2)$, with the number that cross zero before $Q^2 = 15\,$GeV$^2$ being as follows:
$M=9$ -- 92\%;
$M=8$ -- 85\%;
$M=7$ -- 95\%;
$M=5$ -- 97\%;
$M=4$ -- 87\%.
All SPM predictions are listed in Table~\ref{TabZero}.

Referring back to Steps~2\,--\,4, we have checked the influence of using less/more interpolators to define the replica representative.
With $10$ interpolators chosen, the final $M=7$ result for the zero crossing is $10.06_{-1.75}^{+4.62}$GeV$^2$; and with $50$ interpolators, $10.05_{-1.53}^{+3.07}$GeV$^2$.
Plainly, the impact is negligible, producing no change in the central value and having only a modest effect on the uncertainty.
Consequently, all results reported herein are obtained with $25$ interpolators because this value delivers an optimisation of time used versus precision gained.

\begin{figure}[t]
\centerline{%
\includegraphics[clip, width=0.95\linewidth]{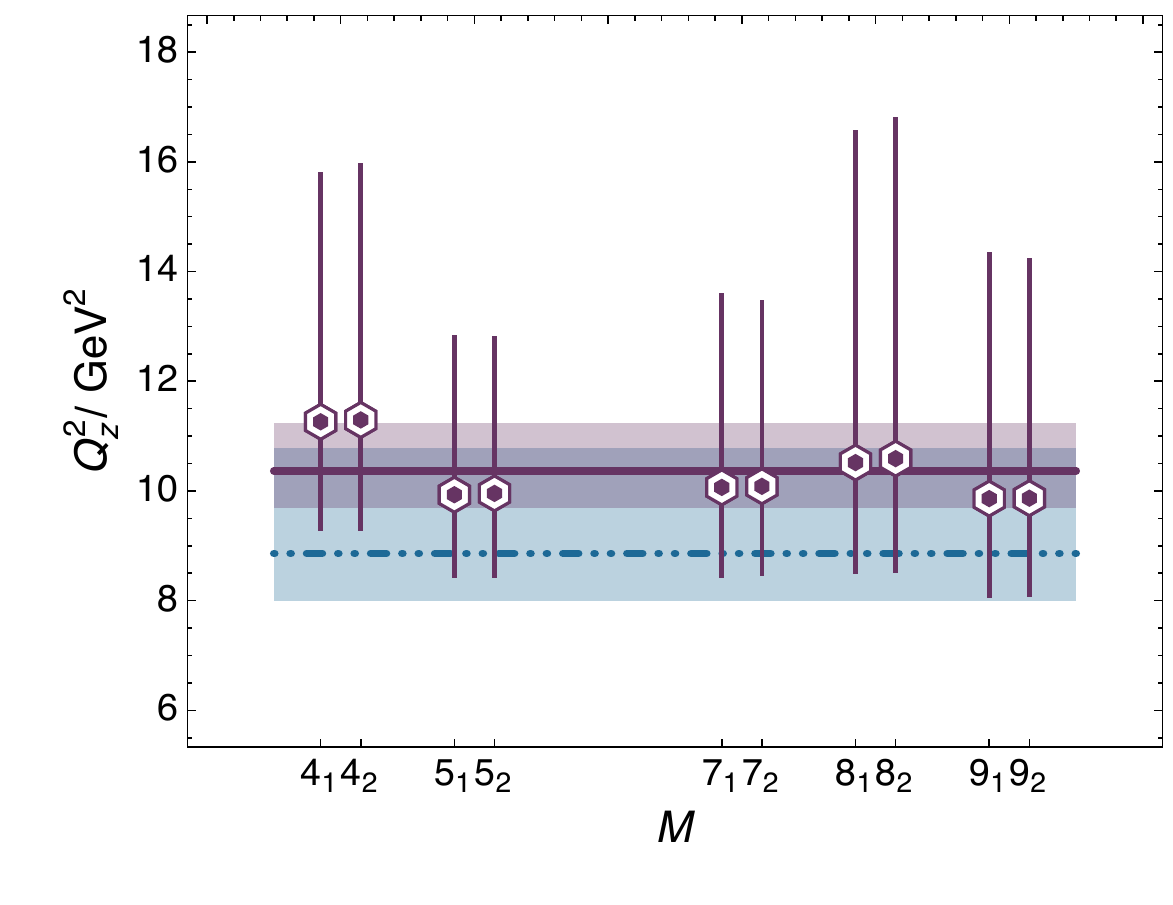}}
\caption{\label{F3}
Points -- $M$-value SPM predictions for the location of a zero in $\mu_p G_E^p(Q^2)/ G_M^p(Q^2)$, drawn from Table~\ref{TabZero}.
Purple line and associated uncertainty band -- combined (final) SPM result in Eq.\,\eqref{GEpzero}.
Blue dot-dashed line and like-coloured uncertainty band -- parameter-free Faddeev equation prediction in Ref.\,\cite{Yao:2024uej}.
}
\end{figure}

Finally, for $M=9$, we repeat Steps 1\,--\,4 with a different set of $15\,000$ randomly chosen initial replicas; and then Steps~2, 4 for $M=8,7,5,4$, in order to demonstrate that the outcomes are independent of input details.
This is plainly the case -- see the results in Table~\ref{TabZero}, which are also drawn in Fig.\,\ref{F3}.

\begin{table}[t]
    \caption{Location of zero in $\mu_p G_E^p(Q^2)/ G_M^p(Q^2)$ predicted by SPM analyses of available data -- see the explanation associated with Steps 1\,--\,4.
    The subscripts $1,2$ indicate the results obtained using two independent, randomly chosen $5\,000$ member replica sets for the ratio data.
    \label{TabZero}}
\begin{center}
\begin{tabular*}
{\hsize}
{
c@{\extracolsep{0ptplus1fil}}
|c@{\extracolsep{0ptplus1fil}}
c@{\extracolsep{0ptplus1fil}}
c@{\extracolsep{0ptplus1fil}}
c@{\extracolsep{0ptplus1fil}}}
$M\ $ & $9_1\ $  & $9_2$ & $8_1\ $ & $8_2\ $ \\
$Q_z^2/{\rm GeV}^2\ $ & $9.86^{+4.45}_{-1.77}\ $  & $9.87^{+4.33}_{-1.75}\ $
& $10.52^{+6.02}_{-1.99}\ $  & $10.59^{+6.19}_{-2.05}\ $  \\[1ex] \hline
$M\ $ & $7_1\ $ & $7_2\ $  & $5_1\ $ & $5_2\ $ \\
$Q_z^2/{\rm GeV}^2\ $
& $10.06^{+3.51}_{-1.60}\ $  & $10.08^{+3.35}_{-1.60}\ $
&$9.93^{+2.87}_{-1.47}\ $  & $9.95^{+2.83}_{-1.50}\ $
\\[1ex] \hline
$M\ $ & $4_1\ $ & $4_2\ $ & & \\
$Q_z^2/{\rm GeV}^2\ $
 & $11.26^{+4.50}_{-1.95}\ $ & $11.29^{+4.64}_{-1.98}\ $  & & \\
\end{tabular*}
\end{center}
\end{table}


It is necessary to supply an explanation for omitting $M=6, 10$ interpolators.
Consider, then, that with $M=2 k + 1$, $k\in \mathbb N$, Eq.\,\eqref{SPMinterpolator} yields interpolating functions that approach a constant at ultraviolet values of $Q^2$.  This constant can take any value -- positive, negative, or zero; hence, no hidden constraint is imposed on the extrapolation.
The situation is different for even values of $M$.
For $M= 4 k$, Eq.\,\eqref{SPMinterpolator} delivers interpolators with odd numerator degree $d=2k -1$ and an even denominator degree, one power larger.
Consequently, the interpolator can be monotonically decreasing and have a single zero crossing at real $Q^2$; so, may be accepted by our algorithm.
On the other hand, for $M=4 k +2$, the numerator has even degree $d=2 k$ and odd denominator degree, again, one power larger.
In each such case, the interpolator must have an even number of zero crossings; and the only monotonic functions satisfying these conditions are those which do not exhibit a zero at finite $Q^2$.
We eliminate all values of $M$ for which this hidden constraint is active.

\section{Final SPM Result}
\label{s3}
All results in Table~\ref{TabZero} are statistically independent and compatible -- the latter observation is highlighted by Fig.\,\ref{F3}.  Hence, they can be averaged.  Care must be taken, however, because the uncertainties are asymmetric.  We follow the method described in Ref.\,\cite{Barlow:2004wg}, sketched hereafter.

Assume that
the ln-likelihood functions associated with the SPM results may be represented by a variable Gaussian form
and the values of the zero crossing, $z$, are distributed around the true value, $\mu$, according to
\begin{equation}
\ln L = -\frac{1}{2} \left( \frac{\mu - z}{\sigma(z)}\right)^2 .
\end{equation}
Supposing that in the neighbourhood of interest, the variation in the standard deviation is linear, \emph{viz}.\
$\sigma(z) = \sigma + \sigma^\prime(z-\mu)$, then one has
\begin{equation}
\ln L = -\frac{1}{2} \left( \frac{\mu - z}{\sigma + \sigma^\prime(z-\mu)}\right)^2 ,
\end{equation}
with $\sigma$, $\sigma^\prime$ determined by requiring that the solutions of $\ln L=-1/2$ are $\mu \pm \sigma_\pm$:
\begin{equation}
\sigma = \frac{2 \sigma_+ \sigma_-}{\sigma_+ + \sigma_-}\,,
\quad
\sigma^\prime = \frac{\sigma_+ - \sigma_-}{\sigma_+ + \sigma_-}\,.
\end{equation}
(With a symmetric uncertainty, one recovers the usual Gaussian distribution.)

\begin{figure}[t]
\centerline{%
\includegraphics[clip, width=0.93\linewidth]{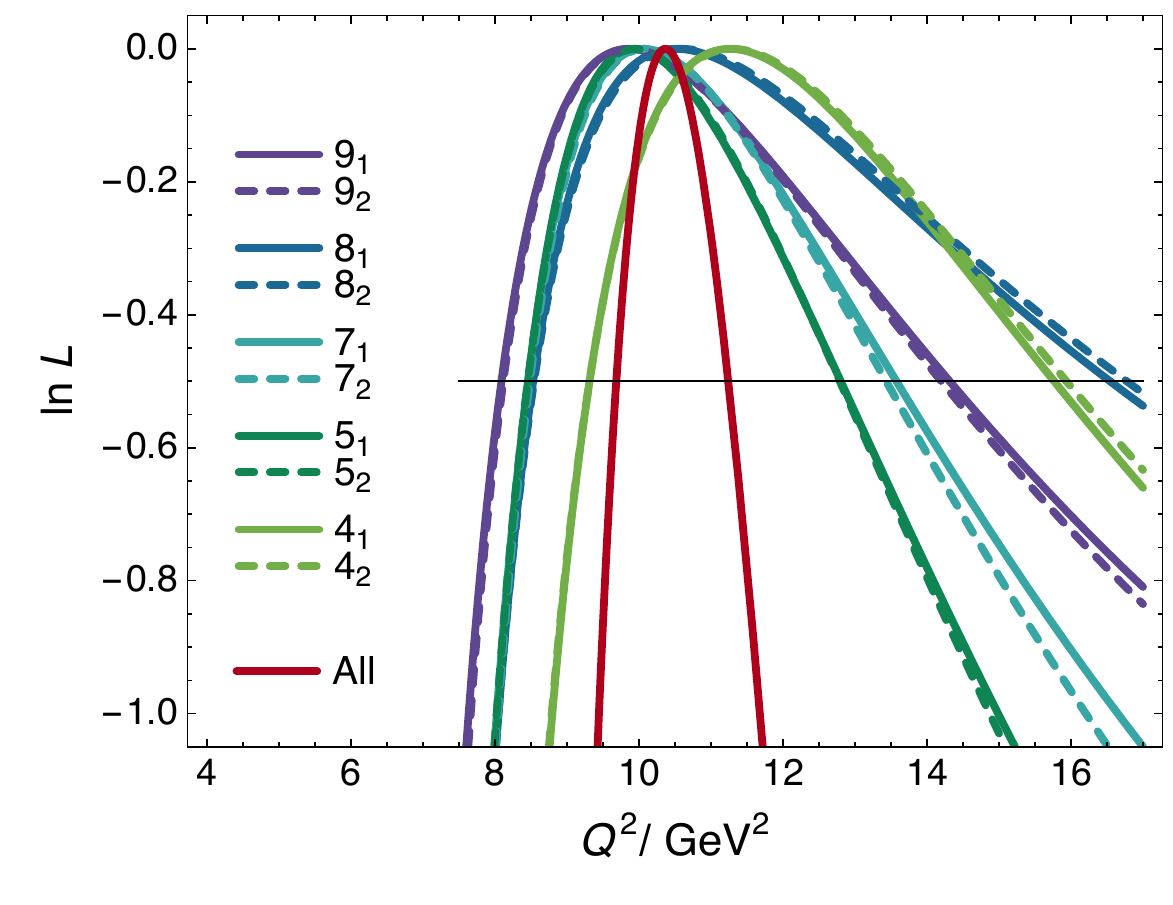}}
\caption{\label{lnlikely}
ln-likelihood functions for the $M$-value SPM data replica analyses listed in Table~\ref{TabZero} along with the combined result obtained as discussed in connection with Eq.\,\eqref{lnlikelycombined}.
}
\end{figure}

In combining $K$ independent, consistent results, one works with the following ln-likelihood function:
\begin{equation}
\ln L_K = -\frac{1}{2} \sum_{i=1}^{K} \left( \frac{\mu_i - z}{\sigma_i + \sigma_i^\prime(z-\mu)}\right)^2
- {\rm max}_{z}\ln L_K\,.
\label{lnlikelycombined}
\end{equation}
Without loss of generality, we have shifted the maximum value to zero.  Now, the $z$ location of the maximum determines the combined value of $\mu$, and $\sigma_\pm$ are the solutions of $\ln L_K = -1/2$.  The individual and combined ln-likelihood functions are displayed in Fig.\,\ref{lnlikely}.  At this point, we can list the final result produced by this function form unbiased SPM analysis of available $R^p(Q^2)$ data, \emph{viz}.\ the SPM predicts a zero in the proton electric form factor at
\begin{equation}
\label{GEpzero}
Q_z^2 = 10.37_{-0.68}^{+0.87} {\rm GeV}^2\,.
\end{equation}
This result is consistent with the parameter-free Faddeev equation prediction in Ref.\,\cite{Yao:2024uej}: $Q_z^2 = 8.86_{-0.86}^{+1.93}\,$GeV$^2$;
and the analyses in Refs.\,\cite{deMelo:2008rj, Cui:2020rmu} also predict zeroes in this neighbourhood.

\begin{figure}[t]
\centerline{%
\includegraphics[clip, width=0.95\linewidth]{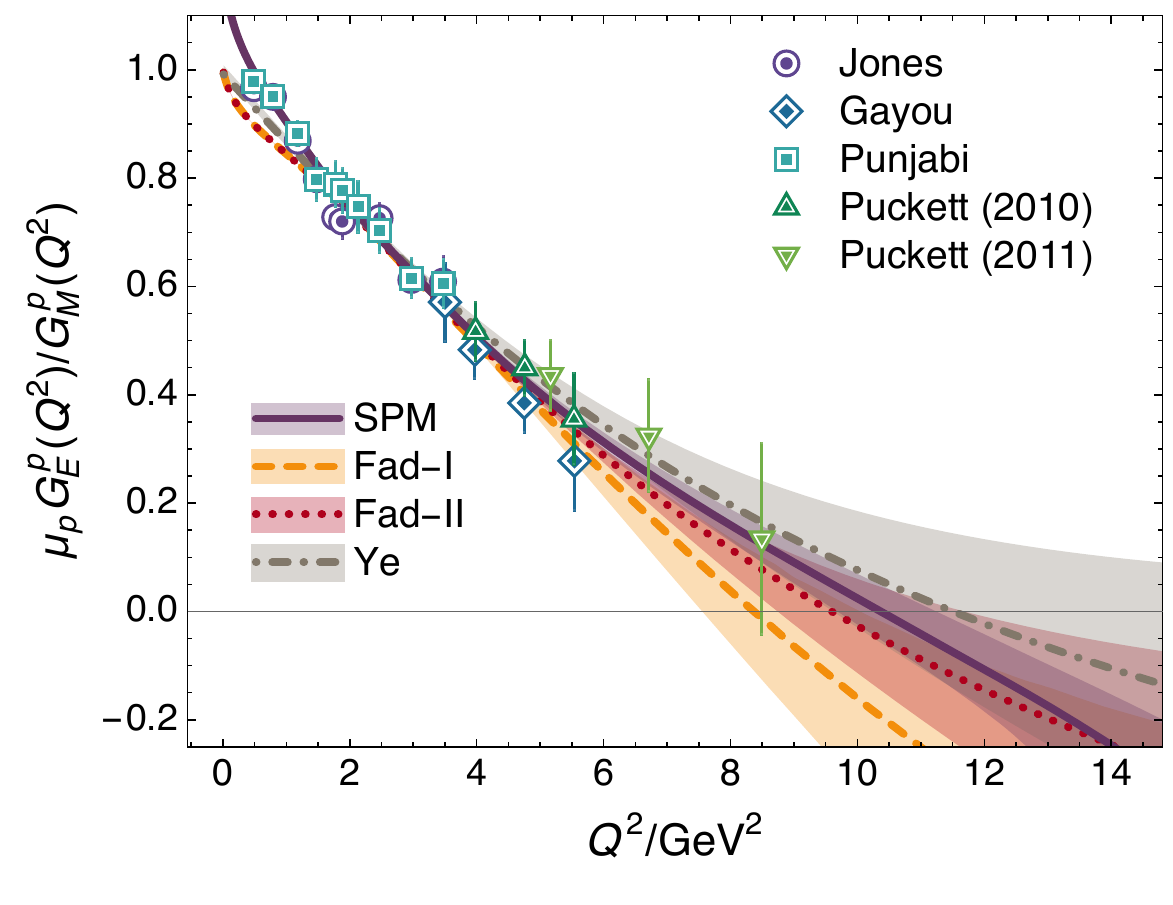}}
\caption{\label{SPMcfYe}
Final SPM prediction for the ratio $\mu_p G_E^p(Q^2)/ G_M^p(Q^2)$.
For comparison, the image also depicts the parameter-free Faddeev equation predictions \cite[Fad-I, Fad-II]{Yao:2024uej} and the result obtained via a subjective phenomenological fit to the world's electron + nucleon scattering data \cite[Ye]{Ye:2017gyb}.
The displayed data are from Refs.\,\cite{Jones:1999rz, Gayou:2001qd, Punjabi:2005wq, Puckett:2010ac, Puckett:2017flj}.
}
\end{figure}

Adding to these comparisons, in Fig.\,\ref{SPMcfYe} we depict our final, objective $Q^2$-dependent SPM result alongside the Faddeev equation predictions from Ref.\,\cite{Yao:2024uej}: the two distinct, yet mutually consistent, Faddeev equation results are obtained using slightly different algorithms for reaching large $Q^2$.  Plainly, even pointwise on $Q^2 \gtrsim 2\,$GeV$^2$, the SPM prediction and Faddeev equation results are consistent within mutual uncertainties.
(The Faddeev equation study omits what may be called meson cloud contributions; hence, deviates somewhat from the data on $Q^2\lesssim 2\,$GeV$^2$ where such effects are noticeable.)

Figure~\ref{SPMcfYe} also includes a result for $R^p(Q^2)$ obtained via a phenomenological function-form-specific practitioner-choice-constrained global-fit to electron + nucleon scattering data \cite{Ye:2017gyb} -- see the grey dot-dashed curve and like-coloured uncertainty band.  In that analysis, a positive-definite proton electric form factor is not excluded, but the favoured result exhibits a zero whose location is consistent with our objective SPM prediction.  Indeed, its pointwise behaviour on $2\lesssim Q^2/{\rm GeV}^2 \lesssim 12\,$ is similar to the SPM result.

\begin{figure}[t]
\centerline{%
\includegraphics[clip, width=0.93\linewidth]{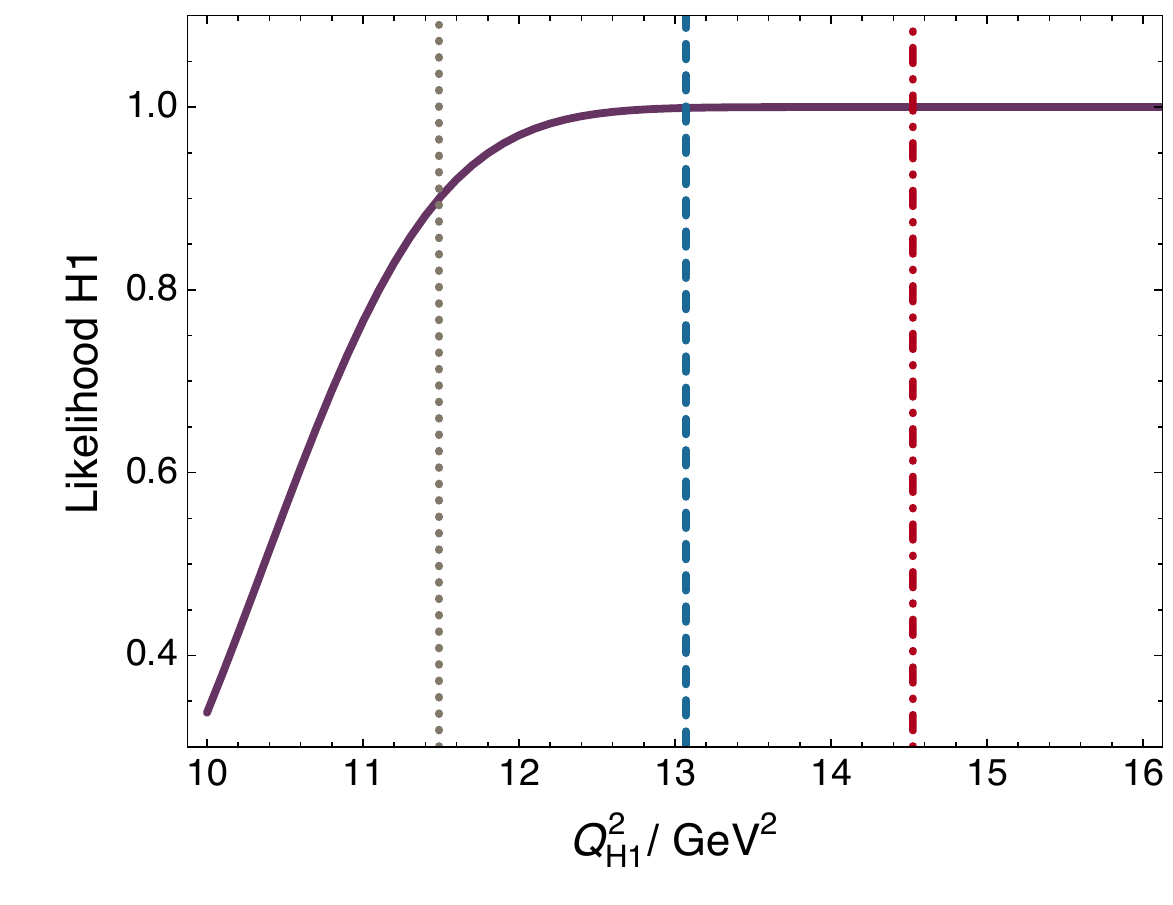}}
\caption{\label{likelyH1}
SPM prediction for likelihood that available data on $\mu_p G_E^p(Q^2)/ G_M^p(Q^2)$ are consistent with a zero in the proton electric form factor on $Q^2< Q_{\rm H1}^2$.
The 90\% and $99.9$\% confidence limits are marked by vertical dotted and dashed lines, respectively.
The boundary associated with H2 is marked by the red dot-dashed line; namely, the likelihood that available data are inconsistent with a zero to the left of this line is $1/1$-million.
}
\end{figure}

It is worth describing the significance of the result in Eq.\,\eqref{GEpzero}.  To that end, consider the following hypothesis:
\begin{description}
\item[H1] The ratio $\mu_p G_E^p(Q^2)/ G_M^p(Q^2)$ exhibits a zero on $Q^2 \leq Q_{\rm H1}^2$.
\end{description}
For any value of $Q_{\rm H1}^2$, the $z$ score is $z=(Q_{\rm H1}^2-Q_z^2)/\sigma_{Q_z}$, where we make the conservative choice $\sigma_{Q_z}=0.87$ because the uncertainty in Eq.\,\eqref{GEpzero} is not very asymmetric.  In Fig.\,\ref{likelyH1} we plot the likelihood, according to our SPM analysis, that the available data is consistent with H1.
With $Q_{\rm H1}^2 = Q_z^2$, the likelihood is 50\%;
with 90\% confidence, one can say the data is consistent with H1 for $Q_{\rm H1}^2 = 11.49\,$GeV$^2$
and with 99.9\% confidence when $Q_{\rm H1}^2 = 13.06\,$GeV$^2$.

It is also worth remarking on a complementary hypothesis; namely,
\begin{description}
\item[H2] The ratio $\mu_p G_E^p(Q^2)/ G_M^p(Q^2)$ is positive definite on $Q^2 \leq 14.49\,{\rm GeV}^2$.
\end{description}
The likelihood that available data are consistent with H2 is $1/1$-million.

\section{Summary and Perspective}
\label{s4}
The possibility that the ratio $\mu_p G_E^p(Q^2)/ G_M^p(Q^2)$, $\mu_p = G_M^p(0)$, possesses a zero at some $Q^2 \gtrsim 6 m_p^2$ has excited great interest for nearly twenty-five years \cite{Jones:1999rz}.  Whether it does and, if so, its location, present fundamental questions to any attempt at solving the strong interaction problem within the Standard Model of particle physics.  The answers will shine light on the character and expressions of emergent hadron mass.

Working with the $29$ available data on $\mu_p G_E^p(Q^2)/ G_M^p(Q^2)$ and without reference to any model or theory of strong interactions, we develop $50\,000$ function-form unbiased continued-fraction interpolator estimates of the curve that underlies the data.
Our analysis scheme -- the Schlessinger point method (SPM) -- is founded in analytic function theory and applicable in the same form to diverse systems and observables.
The SPM returns an objective expression of the information contained in any data under consideration along with a reliable quantitative estimate of interpolation and extrapolation uncertainties.

Our study predicts that, with 50\% confidence, the data are consistent with the existence of a zero in the ratio on $Q^2 \leq 10.37\,$GeV$^2$ [Eq.\,\eqref{GEpzero}].
The level of confidence rises to 99.9\% on $Q^2 \leq 13.06\,$GeV$^2$ [Fig.\,\ref{likelyH1}].
Moreover, the likelihood that the data are consistent with the absence of a zero in the ratio on $Q^2 \leq 14.49\,$GeV$^2$ is $1/1$-million.

New data on the ratio should become available in the foreseeable future.  Meanwhile, the predictions presented herein can serve as useful benchmarks for strong interaction phenomenology and theory.  Some models and phenomenological data fits are challenged by the conclusions that our analysis supports.

\medskip

\noindent\textbf{Acknowledgments}.
%
%
Work supported by:
National Natural Science Foundation of China (grant no.\ 12135007),
Natural Science Foundation of Anhui Province (grant no.\ 2408085QA028),
Spanish Ministry of Science, Innovation and Universities (MICIU grant no.\ PID2022-140440NB-C22),
Project AST22$\_00001\_$X, funded by NextGenerationEU and ``Plan de Recuperaci\'on, Transformaci\'on y Resiliencia y Resiliencia'' (PRTR) from MICIU and Junta de Andaluc{\'{\i}}a;
and completed in part at Institut Pascal, Universit\'e Paris-Saclay, with the support of the program ``Investissements d'avenir'' ANR-11-IDEX-0003-01.

\medskip
\noindent\textbf{Data Availability Statement}. This manuscript has no associated data or the data will not be deposited. [Authors' comment: All information necessary to reproduce the results described herein is contained in the material presented above.]

\medskip
\noindent\textbf{Declaration of Competing Interest}.
The authors declare that they have no known competing financial interests or personal relationships that could have appeared to influence the work reported in this paper.


\end{document}